# Predicting Cyber-Attack using Cyber Situational Awareness: The Case of Independent Power Producers (IPPs)

Akwetey Henry Matey[1], Paul Danquah[2], Godfred Yaw Koi-Akrofi[3]
Department of I.T. Studies, University of Professional Studies, Accra (UPSA), Ghana[1, 3]
Department of I.T. Heritage Christian College, Accra, Ghana[2]

*Abstract*—**The increasing critical dependencies on Internet-of-Things (IoT) have raised security concerns; its application on the critical infrastructures (CIs) for power generation has come under massive cyber-attack over the years. Prior research efforts to understand cybersecurity from Cyber Situational Awareness (CSA) perspective fail to critically consider the various Cyber Situational Awareness (CSA) security vulnerabilities from a human behavioural perspective in line with the CI. This study evaluates CSA elements to predict cyber-attacks in the power generation sector. Data for this research article was collected from IPPs using the survey method. The analysis method was employed through Partial Least Squares Structural Equation Modeling (PLS-SEM) to assess the proposed model. The results revealed negative effects on people and cyber-attack, but significant in predicting cyber-attacks. The study also indicated that information handling is significant and positively influences cyber-attack. The study also reveals no mediation effect between the association of People and Attack and Information and Attack. It could result from an effective cyber security control implemented by the IPPs. Finally, the study also shows no sign of network infrastructure cyber-attack predictions. The reasons could be because managers of IPPs had adequate access policies and security measures in place.**

## I. INTRODUCTION

The massive application of IoT on the electrical grid opened up a huge opportunity to utilize previously untapped processing power to offload custom applications directly to other devices. These transformations are not achievable without experiencing some form of security vulnerabilities on the grid ([1], [2]), even though deploying these business applications on the grid will increase the overall robustness of the grid and reduce communication overhead. A recent attack in 2016 on the Ukrainian power grid was an advanced form of hacking the CI by the Russians extending their intrusion to increase control using the "Crash Override". A related attack occurred in March 2016, compromising the command-and-control (C&C) system on the New York City Dam with a cellular phone. The Stuxnet attack also created awareness of potential cyber threats on power generation companies[3], impacting the grid's reliability [4]. According to [5], various forms of cyber-attack concerning the grid are man-made manipulation. Raikuma, et al [6], have indicated the need to avoid such incidents due to the ripple effects of a power system shutdown. The growing investment of human capital and financial resources injected into CI protection shows the extent to which industry players and the research community understand CI challenges. The increased investment in the sector calls for the need to evaluate cyber situational awareness (CSA) from a human behaviour perspective since the critical infrastructure (CI) falls within a dynamic changing environment. CSA can assist in comprehensively investigating an approach to the ongoing debates relating to cyber security. Cyber awareness for cyber defence generally requires perception, understanding and projection. CSA creates room for predictions in line with an action sequence and effectively plans for new cyber-attacks trends. Hence, the need to identify activities of interest to maintain awareness of a new paradigm in cyber defence. According to Franke and Brynielsson [7], CSA is comparable to insider informants leaking information on an imminent attack. Prior studies had also revealed a range of cyber incidents stemming from minor employee mistakes, misinformation on controls, and highly coordinated, well-planned attack on the critical infrastructure ([1], [8]–[10]). Johnson and Banfield [11] revealed that current cybersecurity defences cannot match the sophistication of embedded technologies attacks capabilities. Hence, the need to evaluate cyber situational awareness (CSA) from human behaviours perspective to predict, detect, and prevent cyber-attack vulnerabilities in a dynamic power generation environment. The concept of cyber situational awareness can be situated based on the insight of individual abilities to distinguish and assess current and future effects in terms of how situations evolve from an attacker's perspective to understand and restrict cyber vulnerabilities [12]. Prior research has made an immense contribution in applying various technologies to support critical cyber incidents. Yet, today's cybersecurity challenges in the power generation sector are increasing and becoming more sophisticated and alarming than we think in the power sector ([5], [13], [14]). The introduction of intelligent information technology equipment such as the Internet of things (IoT) devices and other industrial control systems (ICS) enabled the power generation grid to become more effective and intelligent ([3], [15], [16]). Fig. 1 depicts a visual overview of a possible cyber-attack on the generation and distribution section of the power grid.

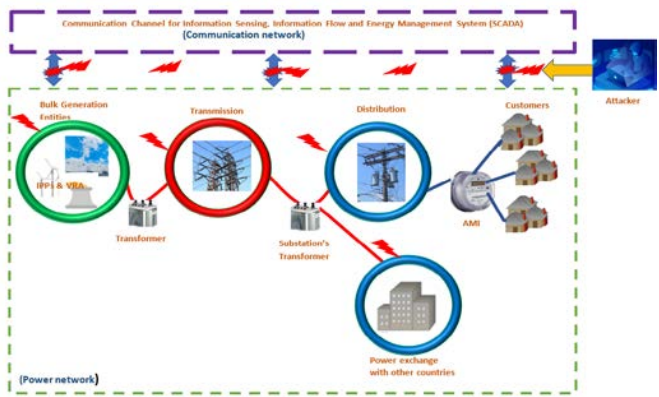

Fig. 1. Smart Grid Perspective in Ghana Source: Authors' Construction (2021).

Based on the above discussion, this current study seeks to establish how CSA elements (Network, Information and People) influence cyber-attacks in the power generation environment in perspective. Digital trust links people, data, and networks [17]. The specific objectives of this study are therefore to:

*1)* Evaluate how staff perceives potential cyber threats in their working environment.

*2)* Evaluate information vulnerabilities and how they influence cyber Attack.

*3)* Assess how cyber situational awareness network vulnerabilities contribute to cyber Attack.

## II. LITERATURE REVIEW

### A. Theoretical Review

Endsley [18] gave three (3) levels of indication to assist in forming a mental model of having a more comprehensive view of an operational environment, as shown in Fig. 2.

In our effort to evaluate cyber situational awareness from human behaviour in an IPPs environment, the authors consider the base of the variable in the conceptual framework in line with the considerations in Table I and Fig. 3.

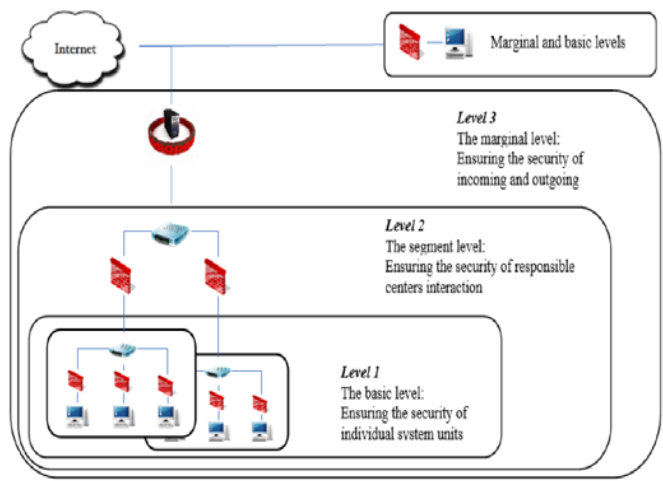

Fig. 2. A Model for Understanding Operational Situation by Endsley.

TABLE I. SHOWS THE MAP OF AUTHORS CONSIDERATIONS AND ENDSLEY

| Focus | Endsley | Authors Considerations |
|---|---|---|
| 1 | Security of individual systems | Security from human behavioural in context interacting with the system of the IPPs |
| 2 | Security of Centre's | Security from the perspective of the IPPs Grid Network |
| 3 | Security concerning incoming and outgoing | protection from human behavioural in context when dealing with sensitive and confidential pieces of information of IPPs |

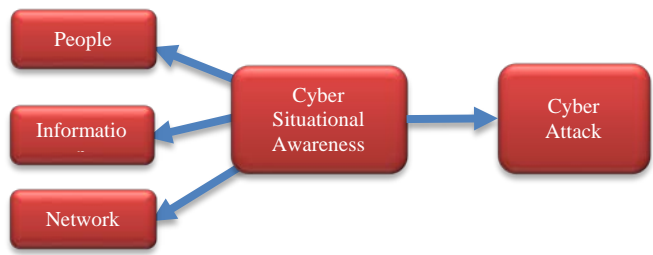

Fig. 3. A Proposed Conceptual Framework Source: Authors' Construction (2021).

The basis for each of the hypotheses used in this study is explained below.

### B. People

In 2019 the Worldwide Threat Assessment by the U.S. indicated that hackers, hacktivists, and insiders pose significant cyber threats to the grid. According to [19], a variation of threat actors pose substantial cybersecurity threats to the electric grid; these actors support grid operations. Ramamurthy and Jain [16] also indicated the difficulty in managing the workforce transformation, which is likely to be the most worrying complication of IoT implementations. Cyberattacks at varying levels of criticality on nations businesses and organizations with an internet presence primarily contribute to human-centric activities [20]. Some persons may work as a team to make decisions and carry out actions [18]. In the work of [21], internal employees working on the grid due to untrained employees or unhappy employees who have hatefulness for other consumers or the service providers also contribute to cyber-attack activities, therefore, seeking to evaluate the vulnerabilities. These factors are considered based on social networking, operating procedure maintenance, and security-related issues, such as user elicitation of information on cyber threats on CI networks, person-to-person interactions challenges concerning user control, and whether or not users strictly adhere to cyber security protocols. Based on the preceding, we propose the following hypothesis.

(H1): Operational staff's cyber security vulnerabilities positively influence cyber-attacks.

(H1.1): Vulnerabilities from operational staff effect of network activities positively influence cyber attacks.

### C. Information

In conceptualizing cyber situational awareness, there is the need to stress the practical concept that conveys security-relevant information that supports the decision-making process [7]. In context, we refer to the logical data flow between network nodes, such as the IoT devices and other intermediary devices, which mandate is to temporarily collect and transmit some form of data emanating from CI activities. A recent report [19] indicated that a critical protection assessment action required to address cybersecurity risks facing the electric grid is Data security and Information protection processes and procedures. Due to power generation system vulnerabilities to cyberattacks against practical state estimation, Zhang [22] develops a comprehensive situational awareness framework for distribution system information on monitoring and controlling state estimation components, cyber-attack detection information, fault location, and voltage control information. Frequently, receiving devices generate information obtained from different devices sources and determine its reliability. Nonconformities between the essential information and possible vulnerabilities areas in the grid can be recognized from emerging information's nature [23]. The constructs objective is to evaluate vulnerabilities and their ramifications on the grid.

- The authors measured how individuals apply cyber security controls when dealing with sensitive and confidential information.

- Whether they had ever experienced information leakages or sensitive information received from the grid comes with inaccuracies.

- If there are restrictions on remote access and finally to enquire if there had been pieces of evidence of frequent misleading information been to receive on their systems.

(H2): The Information handling from operational staff positively influences cyber-attack.

(H2.1): The effect of network activities on information handling influences the cyber Attack.

### D. Network

The communication network and the electrical grid play a significant role in generating and transmitting power to either a sub-station or a customer. Hence, identifying various network vulnerabilities impact assessments will provide knowledge of future impact projection. Prior studies have discussed different techniques to improve cyber situational awareness [7], [24], [25], mainly for analyzing the trends in network traffic. With the evolution of grid networks, there are increasing security threats due to the expanding volume of data transmitted on the grid. Some of these specific cyber incidents on the grid network, as indicated by ([26]–[28]) in their recent study. Because Cyber situation awareness empowers cybersecurity experts to detect and fully understand and anticipate incoming threats, however, our thorough review of the literature revealed no previous studies on how the human behavioural concept can apply in this context. To assess cyber situational awareness of the grid network vulnerabilities and how these vulnerabilities can influence cyber-attacks. Per these constructs, our goal is to evaluate vulnerabilities and their implications for the grid. Concerning the measurement of network vulnerabilities from a human behavioural perspective, authors seek to:

- Enquire whether or not individuals can access the grid network with their devices.

- Find evidence of unauthorized persons accessing the grid network remotely.

- Find the frequency of change of network access policies.

- Enquire if there is evidence of unauthorized IT staff accessing the network remotely.

- Finally, enquire if users can access social media applications on the network.

H3. Negative human behavioural activities on the network positively influence cyber-attacks.

### E. Cyber Situational Awareness

The military is where situational awareness first appears. Situational awareness aims to identify events, causes, consequences, and future projections [18]. It also considers the status and attributes of elements by assessing the present situation and predicting future outcomes based on previous understanding and acquaintance [29]. It becomes possible through the acquisition of data, conception, and synthesis to enable decision-makers to resolve problems with the massive deployment of IoT devices in the power generation sector; for data, acquisition to continuously monitor various sub-systems of the entire power generation system Infrastructure. Therefore, Cyber Situation Awareness (CSA) extends Situation Awareness (SA) to the power generation cyber domain. Hence, we can access fist hand information and seek indications from an attacker's perspective, estimate the impact, to anticipate their actions. Research has carefully refined cyber situational awareness predominantly in the CI ([22], [30], [31]). However, a careful study of the literature did not reveal any prior studies investigating how the Cyber Situation Awareness elements (people, information, and networks) concept can apply in human behaviours from IPPs operational environment. Because according to Michael, et al [32], CSA is the degree to which individuals within a team possess the CSA required in carrying out their responsibilities. We believe cyber vulnerabilities could occur due to various duties discharged by the operational staff of IPPs hence, evaluating human behaviours in the power generation sector. These vulnerabilities can occur in any of the layers; physical layer, information layer, and the human layers in the CI [33]. Also see Appendix.

### F. Cyber-Attack

Cyber-attack issues relating to the smart grid and its impact on the IPPs are increasingly problematic and threatening to a developed and developing economy. In terms of business and human privacy and even national security, the current grid Infrastructure uncertainties manifest due to the power generation sector's Internet of things (IoT). Its cyber security issues have become a significant subject of debate globally. Krishnan, et al [34] indicate evidence of price cap and bid price manipulation and subsequent Attack on the generation unit. Memories of a cyber attack in New York in March 2016 affected the Dam control system with a cellular modem. Such attacks can lead to incapacitating the practical function of the electric grid in line with communication between systems or equipment on the grid network ([8], [15], [35]). It can also harm the effective grid functioning ([1], [36], [37]). Recent studies in the area reveal various cyber-attacks such as False data injection attacks ([27], [38], [39]), Denial-of-service attacks identified by ([40], [41]), Distributed Denial of Service attack indicated by [42], Man-in-the-middle attack [6], Malware Attack by ([43], [44]), State Estimation attack [45], and Price manipulation and Misrepresentation of values attack[1] Coordinated ([10], [43]), etc.

### G. Related Work

The new paradigm of human-centric warfare on cybersecurity is wealth looking into because insider threats can be disruptive and equally malicious as an attack from outside an organization [20]. Cyber situational awareness gives insight into understanding an impending phenomenon [46]. Since many sectors in the economy are primarily dependent on Critical infrastructures (CI) [11], where information is sent and received for processing to predict possible future threats and adequately plan the power generation environment. According to Nekha and Dorosh [23], the critical aspect of cyber situational awareness can timely deal with an emerging threat model. Hence, critically examine elements such as people, information, and networks in the context of cyber situational awareness within the operational environment of the independent power producers (IPPs) in perspective. The growing adoption of high IoT devices connected to the Internet and the use of the global positioning system to harmonize grid operations contributes to grid vulnerabilities [19]. Current studies in the sectors focus on the cyber-physical aspect of the grid; [47] examine the multi-stage attacks using WannaCry ransomware. The study of [48] accesses the vulnerability with the local power trading. Sharafeev et al [1] develop an algorithm to monitor electrical power systems (EPS) cyber-attacks. A framework for assessing critical infrastructure in an attack was created by Akhtar, et al [49]. Sarangan, et al [50] Analyzed cyber-attacks of the power grid considering the ill-effects of increasing renewable penetration [39]. Develop a systematic two-stage approach for detecting false data injection (FDI). Roy and Debbarma [27] also Proposed a cyber-attack detection and mitigation platform which uses forecasted data. Raikuma, et al [6] also demonstrate the impact of the man-in-the-middle attack, which exploits vulnerabilities in the Generic Object-Oriented Substation Event (GOOSE). Prior related research work was sort of giving indications of lack of effort to explore vulnerabilities from Human Behavioral perspective particularly, in the power generation sector. Hence our current study seeks to critically evaluate and access the cyber vulnerabilities within the IPPs using cyber situational awareness.

Based on the above reviews, we propose five (5) hypotheses, as follows:

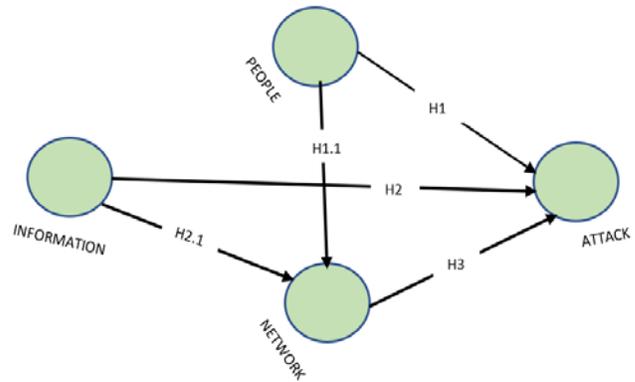

Fig. 4. Conceptual Framework.

Based on the above research hypotheses, the influence of cyber situational awareness on People, Information, and Network cyberattacks is present in Fig. 4.

## III. RESEARCH DESIGN AND METHODOLOGY

### A. Research Design

The approach of our current study is purely quantitative. We sort to use this approach because it involves an empirical investigation of the social phenomena of Cyber Attacks from Human Behavioral in the context of the IPPs. Therefore, the employs quantitative data using the survey method to collect and gather the information required for the analysis.

### B. Data Collection

Questionnaires were administered purposively to a sample of selected units of the IPPs staff who are directly involved in the organization's day-to-day grid operations from March to May 2021. We sort to use questionnaire because of anonymity among respondents and its low-cost implication. The questionnaire consists of four (4) parts: Demographic factor, People, Information, and Network. The questionnaire design is founded on [51] conceptual framework using a Likert scale from (One)1 for strongly disagree to (Five) 5 for strongly agree.

### C. Population and Sampling Procedures

The population studied is GRIDco and five (5) IPPs in Ghana that is actively operating with GRIDco in the power generation sector to assess Cyber Attacks on the electrical grid from human behaviour in context. Our estimated sample frame is 300, based on the various units within the IPPs: Supervisory Control and Data Acquisition (SCADA) Unit, Telecommunication Unit, and the Management Information Systems (MIS). The sampling method employed was probability sampling (random sampling). Choice of employees was at random within the various IPPs units selected to answer the questions

### D. Sample Size Calculation

Our Sample size is calculated based on Yamane 1967 formula with 95 per cent confidence level plus or minus 5 per cent confidence intervals using the formula n = $N/1+N(e)^2$ where **n=** is the sample size N =is the population, and **e**= is the error margin. Although a sample size of 171 is obtained per our calculation from the estimated population of 300. Meanwhile, the Actual sample size used for the analysis is based on the number of respondents, which is (238) was used for the study because, with most research, a large sample size gives more reliable results than smaller samples.

## IV. ANALYSIS AND RESULTS

The PLS-SEM approach seeks to support evaluating patterns of causality of target constructs in the structural model. It also provides more detailed statistical associations supporting variables included in a model. Authors' employees' PLS-SEM in the study since it allows relatively more minor samples than other statistical software like Amos and Lisrel. PLS-SEM does not enforce stringent assumptions on data distribution [52]. 238 is the valid responses received, which forms the basis of data analysis; 86.1% were males, while 13.9% were from females (see Table 2) on demographic of respondents. Using the two-step approach to evaluating structural equation models as suggested by [53], we began examining the measurement model to assess the instrument's reliability and validity. We then looked at the structural model based on the hypotheses proposed in this study.

### A. Measurement Model

Following our measurement model assessment, we identified four(4) items (AK2, C5, C6, and C15) is removed from the study as a result of their low loadings, which is less than 0.600as recommended by ([54], [55]). In assessing the reliability of the constructs for our measurement Models, we used Cronbach's alpha and composite reliability measures to test the model's internal consistency. Hence Cronbach's alpha and composite reliability for each construct are adequate, as indicated in Table III. Composite reliability should be higher than 0.6 and 0.70 ([56]–[58]). With indicator factor loadings surpassing 0.5, Nunnally [58] and Hair, et al [59] recommended that the Average Variance Extracted for each variable should exceed 0.5 [56] to assess convergent validity [59]. Indications of convergent instrument validity are present in Table III. Hence, convergent validity is suitable since the AVE values exceed 0.500 [60]. We strictly adhere to the Fornell–Larcker criterion in reporting the discriminant validity, which posits that AVE for each latent construct should be higher than the construct's highest squared correlation with any other latent construct [56].

Additionally, the loadings of each indicator should be greater than all its cross-loadings [53]. An inspection of indicator cross-loadings in Table III shows that all indicators load their highest on their respective construct. No Indicator loads higher on other constructs than on its intended construct. Evident in Table 4 shows that the square root of the AVEs for each construct is greater than the cross-correlation for different constructs. Based on these results is the established discriminant validity of the instrument. We also measured Discriminant validity using the Heterotrait-Monotrait Ratio of correlation(HTMT) criterion by [61], [62], see Table V below.

TABLE II.    DEMOGRAPHY OF RESPONDENTS

| ATTRIBUTE | CATEGORIES | PERCENTAGE (%) |
|---|---|---|
| Gender | Male | 86.1 |
|  | Female | 13.9 |
| Age | 18 -30 | 0.8 |
|  | 31-40 | 12.6 |
|  | 41-50 | 19.7 |
|  | 51-60 | 37.8 |
|  | 61+ | 29 |
| Education | Diploma | - |
|  | HND | - |
|  | First degree | 23.1 |
|  | Postgraduate degree | - |
|  | Professional qualification | 35.7 |
|  | Masters | 41.7 |
|  | PhD | - |
| Years of experience | 1-5 | 1.7 |
|  | 6-10 | 15.5 |
|  | 11-15 | 32.8 |
|  | 16-20 | 28.2 |
|  | 21 and above | 21.8 |

TABLE III.    LOADINGS OF RELIABILITY AND VALIDITY RESULTS

|  | ATTACK | INFORMATION | NETWORK | PEOPLE | CA | CR | AVE |
|---|---|---|---|---|---|---|---|
| AK1 | **0.851** | 0.362 | 0.352 | -0.287 | **0.875** | **0.921** | **0.797** |
| AK3 | **0.896** | 0.233 | 0.215 | -0.509 |  |  |  |
| AK4 | **0.929** | 0.564 | 0.568 | -0.454 |  |  |  |
| C1 | -0.436 | -0.033 | -0.058 | **0.948** | **0.934** | **0.953** | **0.836** |
| C2 | -0.466 | -0.27 | -0.259 | **0.894** |  |  |  |
| C3 | -0.163 | 0.386 | 0.394 | **0.742** |  |  |  |
| C4 | -0.447 | -0.014 | -0.013 | **0.938** |  |  |  |
| C7 | 0.394 | **0.908** | 0.849 | -0.051 | **0.922** | **0.945** | **0.811** |
| C8 | 0.392 | **0.856** | 0.839 | -0.17 |  |  |  |
| C9 | 0.413 | **0.919** | 0.827 | -0.02 |  |  |  |
| C10 | 0.459 | **0.971** | 0.949 | -0.103 |  |  |  |
| C11 | 0.417 | 0.872 | **0.909** | -0.166 | **0.916** | **0.934** | **0.782** |
| C12 | 0.338 | 0.732 | **0.858** | -0.081 |  |  |  |
| C13 | 0.422 | 0.91 | **0.887** | -0.03 |  |  |  |
| C14 | 0.421 | 0.887 | **0.947** | -0.076 |  |  |  |

TABLE IV.    DISCRIMINANT VALIDITY USING FORNELL–LARCKER CRITERION

| | ATTACK | INFORMATION | NETWORK | PEOPLE |
|---|---|---|---|---|
| ATTACK | 0.892 | | | |
| INFORMATION | 0.454 | 0.914 | | |
| NETWORK | 0.446 | 0.949 | 0.901 | |
| PEOPLE | -0.475 | -0.094 | -0.097 | 0.885 |
| | | | | |

Notes: Construct correlations with the square root of AVE along the diagonals

TABLE V.    HETEROTRAIT-MONOTRAIT RATIO HTMT

| | ATTACK | INFORMATION | NETWORK | PEOPLE |
|---|---|---|---|---|
| ATTACK | | | | |
| INFORMATION | 0.477 | | | |
| NETWORK | 0.469 | 1.017 | | |
| PEOPLE | 0.47 | 0.241 | 0.247 | |

*B. Structural Model*

In this section, overall explanatory power, Amount of variance explained by the independent variables, the degree of strength of each path is assessed. To estimate path significance, we applied bootstrap. We also assess the quality of the structural model with the coefficient of determination ($R^2$) and standardized root mean square residual (SRMR) ([63] [64]). Our structural model results are present in Fig. 5 and Table 6. Regarding our (H1), The result shows even though people have a negative influence in predicting cyber-attack plays a significant role for an attack to occur (β = -0.435; p =0.000). (H1.1) However, the mediation effect of network activities from people to attack does not have a negative relationship and plays no significant role in predicting cyber-attack within the grid. (β = -0.008; p =0.643). (H2) our results reveal that information handling is significant and positively influences cyber-attack predictions within the grid. (β = 0.296; p=0.020). Meanwhile, (H2.1) information handling positively affects network activities within the grid and significantly influences cyber-attacks (β = -0.949; p =0.000). (H3) was found to have network activities positively affect predicting cyber-attack but not significant (β = 0.123; p =0.300). Hence in this study, we did not find support for H3 and H1.1, as indicated in table 6 below. According to [62], the predictive validity of variance is a criterion for determining a model's prediction accuracy. Hence the coefficient of determination ($R^2$) is the output of regression value as variance proportion in endogenous variable predicted by exogenous variable.

$R^2$ values range from 0 to 1; A higher value is said to have a higher level of $R^2$ of .75 is substantial, .50 is moderate, and .25 is considered as weak ([65], [66]).

This study shows Attack (0.396, being Moderate) and Network with (0.901, substantial) value. In conclusion, the $R^2$ indicates a sufficient level of $R^2$ (see Table VI) and Fig. 5.

The authors performed analysis to assess the mediation role of the Network between People and Attack. The study results in (Table VII) reveal that the total effect of People on Attacks, even though negative is significant (H1.1: β = -0.436, p= 0.000). With the introduction of the mediator variable Network, the impact of People on Attack gives negative effect but significant (β = -0435, p=0.000). The indirect impact of people on Attacks through networks is insignificant (β = 0.116, p=0.318), which indicates no mediations effect between the association of People and attacks.

Finally, assessing the mediation role of the Network between Information and Attack. Our results in (Table VII) also show that the total effect of Information on Attacks has a positive impact and is significant (H2.1: β = 0.413, p=0.000). By introducing the mediator variable Network, the information effect on the Attack shows a positive impact and is significant (β = 0.296, p=0.022). The indirect implications of Information on Attacks through networks have a positive effect and are insignificant (β = -0.001, p= 0.768), which indicates no mediations effect between the association of Information and Attack.

TABLE VI.    PATH COEFFICIENTS AND THEIR SIGNIFICANCE

| Hypotheses | Path | Standardized path coefficient | T Statistics | P Values | Result |
|---|---|---|---|---|---|
| H1 | PEOPLE -> ATTACK | -0.435 | 5.704 | 0.000 | Supported |
| H1.1 | PEOPLE -> NETWORK | -0.008 | 0.463 | 0.643 | Not Supported |
| H2 | INFORMATION -> ATTACK | 0.296 | 2.330 | 0.020 | Supported |
| H2.1 | INFORMATION -> NETWORK | 0.949 | 91.394 | 0.000 | Supported |
| H3 | NETWORK -> ATTACK | 0.123 | 1.037 | 0.300 | Not Supported |
| | | $R^2$ | | | |
| | ATTACK | 0.396 | | | |
| | NETWORK | 0.901 | | | |

Notes: SRMR= 0.155; ns-not significant.

TABLE VII.    MEDIATION RESULTS

| | TOTAL EFFECT | | DIRECT EFFECT | | | INDIRECT EFFECT | |
|---|---|---|---|---|---|---|---|
| | PATH COEFFICIENT | P-VALUES | PATH COEFFICIENT | P-VALUES | | PATH COEFFICIENT | P-VALUES |
| PEOPLE -> ATTACK | -0.436 | 0.000 | -0.435 | 0.000 | PEOPLE -> NETWORK-> ATTACK | 0.116 | 0.999 |
| INFORMATION -> ATTACK | 0.413 | 0.000 | 0.296 | 0.022 | INFORMATION -> NETWORK-> ATTACK | -0.001 | 0.296 |

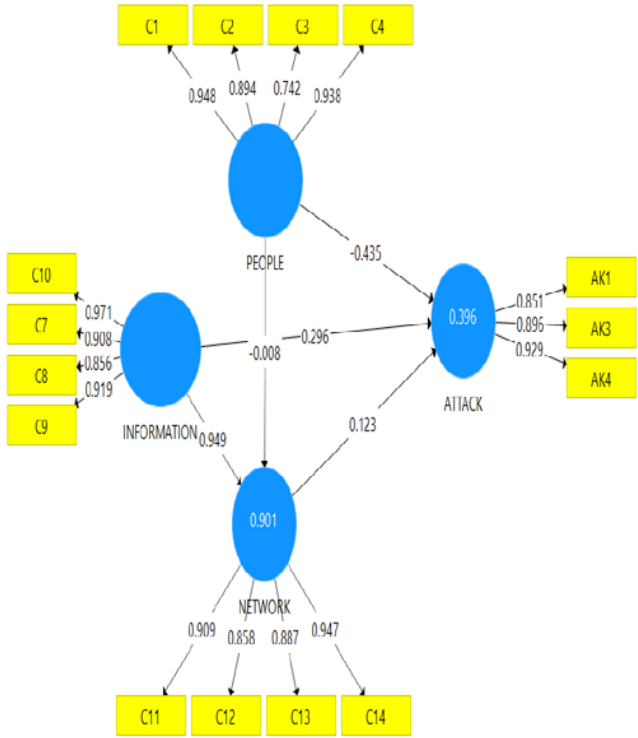

Fig. 5.    PLS Results for Structural Model.

## C. Effect Size ($f^2$)

Effect size is a concept to measure how strong the relationship of an indicator or the effect of exogenous constructs to endogenous constructs is. It examines changes in $R^2$ value when an exogenous construct is detached from the model. An effect size of.02 has a minor influence, a value of.15 has a medium effect, and a value of.35 has a significant impact. [66]. Hence the current study revealed four correlational effect sizes. C15 with a value of 0.018 as a small effect, AK2 and C6 also gave indications of values of 0.264 and 0.543, respectively, having an effect size of the medium impact.C5 recorded the most significant effect sizes with the value of 0.617.

## V. IMPLICATIONS OF THE STUDY

CSA aims to improve the quality of appropriateness of concerted decision-making concerning the protection of the electrical grid. Invalidating our hypothesis, although (H1) posited there is a negative effect of people in predicting cyber-attack, it also plays a significant role for an attack to occur in the operational environment of the IPPs. Therefore, consider cyber security measures to ensure operational staff knows potential external cyber threats. Ensuring that the grid and its sub-systems are fully protected, prohibiting personal devices from accessing and transmitting information on the grid. Finally, steps must be taken to monitor users' grid activities strictly. Hypothesis (H1.1) indicates no mediation effect between the association of People and attacks. Explaining that, irrespective of users' activities such as personal devices, remote access to the network by I.T and Non-I.T staff can prevent an attack on the electrical grid network with the proper security controls. Frequent changes in network access security policies (H2) suggest that information handling is significant and positively influences cyber-attack. Therefore, lack of CSA from the perspective of information accessibility and distribution concerning cyber security controls will distort confidential and sensitive information; hence the need to ensure the information on the grid activities is well-coordinated and accessible to only specific users within the operational environment. The mediation effect of H2.1 indicates no mediations effect between the association of Information and Attack. Reasons could be due to an effective cyber security control is implemented by the IPPs. Information from the grid received with inaccuracies is attributed to frequent updates of security policies in the context of perspective network access. (H3) reveals no sign of cyber-attack predictions. It indicates that the grid infrastructure network managers have put adequate security measures. Such as frequent network access policies, hence irrespective of who performs activities on the network do not translate into any negative impact on the grid network.

## VI. CONCLUSION

Over the years, the power generation and distribution sectors have seen an increase in the number of independent power producers in the electricity market. This sector of power generation has advanced its operations with the intervention of internet of things (IoT) technology and other electronic equipment making the entire grid network susceptible to attack [1], [67], [68]. From the literature perspective, very little is known about user behaviour concerning cyber vulnerabilities in the sector. Bradley [69] insider threats are the most expensive threats challenging to address threats to people, information, and technology in the business environment. In this regard, our research helps to improve the understanding of user behaviour in the context of Ghana's electricity generation sector. From the context of cyber situational awareness (CSA), authors sort to (1) evaluate how the operational staff of IPPs perceive cyber vulnerabilities within their operational environment, (2) evaluate the information vulnerabilities that seeks to influence cyber-attack and finally (3) assess network vulnerabilities that contribute to cyber-attack on the electrical grid. Our findings show that People's construct negatively predicts cyber-attack but plays a significant role in attacking the electrical grid. The authors also realized no mediation effect from people and attack network activities, probably because of constant security controls measures such as frequent update network access policies. Meanwhile, authors also realized that information handling is significant and positively influences cyber-attack, which calls for well-coordinated cyber security controls on the grid activities in line with confidential and sensitive information in the operational environment of IPPs. In addition, there was no indication of mediations effect from network activities between the association of Information and Attack. Such development can be to adequate security measures being put in place to ensure information from the grid are received devoid of errors. Finally, managers of IPPs infrastructure networks seem to have suitable security measures concerning the network activities. Hence irrespective of the numerous activities on the network cannot easily translate into any negative cyber-attack impact on the grid network.

APPENDIX 1

Instrument: Cyber Situational Awareness(Csa)

SECTION A: DEMOGRAPHIC (Please tick appropriately)

1) Please indicate your gender:

   Male [ ]

   Female [ ]

2) Please indicate your age category:

   18 -30 [ ]

   31-40 [ ]

   41-50 [ ]

   51-60 [ ]

   61 [ ]

3) Please indicate your highest category:

Diploma **[ ]**

HND  [ ]

First degree [ ]

Postgraduate degree [ ]

Professional qualification [ ]

Masters  [ ]

PhD  [ ]

4) Please indicate your years of experience on the job:

1-5 [ ] 6-10 [ ] 11-15 [ ] 16-20 [ ] 21 and above [ ]

SECTION B: Please tick appropriately

| Please tick the correct numeric response to each question | | 1= Strongly Disagree, 2=Disagree, 3=Neutral, 4= Agree, 5= Strongly Agree | | | | |
|---|---|---|---|---|---|---|
| | **People** | 1 | 2 | 3 | 4 | 5 |
| C1 | I am aware of potential cyber threats from external sources on our network | | | | | |
| C2 | I am aware that our systems are well secured | | | | | |
| C3 | I know my colleagues use their devices on our network | | | | | |
| C4 | It is difficult to control other computer users on the network. | | | | | |
| C5 | my colleagues and I always strictly follow our cyber security protocols | | | | | |
| | **Information** | | | | | |
| C6 | We do not apply cyber security controls when interacting with confidential and sensitive information | | | | | |
| C7 | We usually experience leakages of vital and sensitive information | | | | | |
| C8 | I have access to data remotely without restriction | | | | | |
| C9 | The Information sometimes receive from the grid comes with inaccuracies due to cyber Attack | | | | | |
| C10 | We frequently receive misleading information from our systems | | | | | |
| | **Network** | **1** | **2** | **3** | **4** | **5** |
| C11 | I can access the network with my devices | | | | | |
| C12 | I am aware staff can access social media applications on the network | | | | | |
| C13 | I am aware unauthorized non-IT staff can access the network remotely. | | | | | |
| C14 | I am aware unauthorized IT staff can access the network remotely. | | | | | |
| C15 | Network access policies changes frequently | | | | | |
| | **Attack** | **1** | **2** | **3** | **4** | **5** |
| AK1 | There is the manipulation of other components on the grid by attackers | | | | | |
| AK2 | Attackers take advantage of cyber vulnerabilities in the IoT devices | | | | | |
| AK3 | We experience loss of generations capability disabling power generation network by attackers | | | | | |
| AK4 | I have access to all systems resource | | | | | |